\documentclass[aps,prl,twocolumn,showpacs]{revtex4}

\usepackage{graphicx}
\usepackage{amsmath}
\usepackage{amssymb}

\begin{document}

\title{
Conformational Temperature Characterizing the Folding of a Protein
}

\author{Naoko Nakagawa}
\affiliation{
College of Science, 
Ibaraki University,
Mito, Ibaraki 310-8512, Japan}

\begin{abstract}
The time sequences of the molecular dynamics simulation for 
the folding process of a protein is analyzed with the 
inherent structure landscape
which focuses on configurational dynamics of the system.
Time dependent energy and entropy for inherent structures are introduced
and from these quantities a conformational temperature is defined.
The conformational temperature follows the time evolution of a slow relaxation
process and reaches the bath temperature when the system is equilibrated.
We show that the nonequilibrium system is described 
by two temperatures, 
one for fast vibration and the other for slow configurational 
relaxation, while the equilibrium system is by one temperature.
The proposed formalism is applicable widely for the systems with 
many metastable states.
\end{abstract}

\date{March 21, 2007}

\pacs{
05.70.Ln  
87.15.Cc  
05.90.+m  
}
\maketitle

Slow relaxations of non-equilibrium systems are a very general process
which are typically observed in glasses \cite{Houches2002}.
For various slow relaxations a major interest is how the relaxation proceeds, 
and inherent structures, i.e.\
the local minima of their energy landscape, are powerful to investigate it 
from the point of energy landscape.
Their statistics might provide thermodynamic aspects in system's configuration
not only at equilibrium but at non-equilibrium 
\cite{Stillinger_Weber,Sciortino,Crisanti-Ritort,Nieuwenhuizen}.

The folding of proteins
provides a nice example of slow relaxation in an out-of-equilibrium complex physical system 
without homogeneity nor extensivity.
It evolves from random coil states to their
own native conformation in $\mu$s or $m$s,
while their vibrational dynamics takes place in time scales of the
order of $f$s or $p$s. 
Proteins can be investigated in the
spirit of glasses by characterizing their inherent structures.
We showed that the density of states in the Inherent Structure Landscape
(ISL), i.e. the landscape of inherent structure energies, 
of a model protein depends exponentially on their energy
\cite{Nakagawa_Peyrard}, so that the model has some
similarities with the simple models of glasses such as spin
glasses \cite{Bray}, Lennard-Jones systems \cite{Doye,Sciortino-Kob}
 and the trap model \cite{Fielding},
despite of its nonextensivity and heterogeneity.
This letter shows that the ISL is not only useful to study the
equilibrium properties of a protein 
but can
also be used in practice to characterize its slow evolution towards the
folded configuration
 by proposing a conformational temperature {\it that follows
the time evolution of a slowly relaxing non-equilibrium
system}.

\paragraph*{Model: }
We consider an off-lattice G\=o protein model with a slight frustration 
\cite{Nakagawa_Peyrard}.
The protein is reduced to its backbone without side chains,
each residue being represented by its $C_{\alpha}$ atom.
The effective potential between the $C_{\alpha}$ elements in the backbone 
is designed from the experimentally determined native structure.
We study proteinG which has 56 residues \cite{PDB}.
At low temperature, the model protein folds to the native structure,
while at high temperature it
denatures.  Folding-unfolding transition occurs at $T_f$, and
around $T_d \approx 0.4 T_f$ a dynamical transition is observed
below which the fluctuations of the protein structure are dramatically
reduced. 

In numerical investigation, 
molecular dynamics simulations with Langevin equations are applied,
where a denatured initial condition is
prepared by a simulation at $1.7\; T_f$ at which the protein is
in the random coil state.  A sudden cooling to
$T < T_f$ is applied at $t=0$.  
For each value of $T$, $200$ initial conditions are sampled.

\paragraph*{Definitions of quantities: }

Slow dynamics comes from itinerant transitions 
among local energy well
while the fast dynamics comes from rapid vibrations inside each
well.  
Such slow dynamics can be
characterized by ``inherent structures'' \cite{Stillinger_Weber},
in which the set of conformations belonging to the
same basin labeled by the index $\alpha$
are mapped to its local energy minimum 
with energy $e_{\alpha}$.

Let $w(\Gamma, t,T)$ be the probability density for the system to be
at point
$\Gamma$, up to $d \Gamma$,
in phase space at time $t$ and let $\Gamma_{\alpha}$ be the
region 
corresponding to a basin $\alpha$. The
probability weight to observe the basin $\alpha$ is 
\begin{equation}
w_{\alpha}(t,T)=\int_{\Gamma_{\alpha}} w(\Gamma, t,T) d\Gamma,
\label{eqn:weight}
\end{equation}
where $\sum_{\alpha} w_{\alpha}(t,T)=1$.
The probability weight $w_{\mathrm{v}}^{\alpha}(\Gamma, t,T)$ 
normalized in the basin $\alpha$ is introduced as
\begin{equation}
w(\Gamma, t,T)=w_{\alpha}(t,T)\cdot w_{\mathrm{v}}^{\alpha}(\Gamma, t,T).
\end{equation}
To be consistent with Eq.(\ref{eqn:weight}),
$\int_{\Gamma_{\alpha}} w_{\mathrm{v}}^{\alpha}(\Gamma, t,T) d\Gamma=1$.

For each point $\Gamma$, the value of the energy
$e(\Gamma)$ is determined from the model potential.  When the point
$\Gamma$ belongs to basin $\alpha$, the energy can be rewritten as
$e(\Gamma)=e_{\alpha}+\Delta V_{\alpha}(\Gamma)$, where $\Delta
V_{\alpha}(\Gamma)$ is the potential at
point  $\Gamma$ measured
from the minimum $e_{\alpha}$.  The mean energy $U$ at time $t$ is
given by
\begin{eqnarray}
&U(t,T)&\equiv\int e(\Gamma) w(\Gamma, t,T) d\Gamma \nonumber\\
&=& U_{\mathrm{IS}}(t,T)+\sum_{\alpha} U_{\mathrm{v}}^{\alpha}(t,T) w_{\alpha}(t,T).
\label{eqn:Uis}
\end{eqnarray}
This expression splits the energy into two components, 
$U_{\mathrm{IS}}(t,T)\equiv \sum_{\alpha} e_{\alpha} w_{\alpha}(t,T)$ 
and $U_{\mathrm{v}}^{\alpha}(t,T)\equiv\int_{\Gamma_{\alpha}}\!\!\!\!\Delta V_{\alpha}(\Gamma)w_{\mathrm{v}}^{\alpha}(\Gamma, t,T) d\Gamma$.
$U_{\mathrm{IS}}(t,T)$ is the mean inherent structure energy at time $t$
which reflects slow structural changes due to the transition among the
basins while
$U_{\mathrm{v}}^{\alpha}(t,T)$ is 
the mean vibrational energy inside basin $\alpha$.
This split does not require the separation of time scales between the two modes, 
but numerical simulations below deal with separated situations.

Since the inherent structure is discrete, 
the entropy for the inherent structures can be defined as
\begin{equation}
S_{\mathrm{IS}}(t,T)\equiv -k_{\mathrm{B}} \sum_{\alpha} w_{\alpha}(t,T) \log w_{\alpha}(t,T).
\label{eqn:Shanon}
\end{equation}
The total entropy $S(t,T)$
splits into $S_{\mathrm{IS}}(t,T)$ and a vibrational contribution
$S_{\mathrm{v}}^{\alpha}(t,T)$
according to
\begin{eqnarray}
S(t,T)&\equiv& -k_{\mathrm{B}} \int w(\Gamma, t,T) \log w(\Gamma, t,T) d \Gamma\nonumber\\
&=&S_{\mathrm{IS}}(t,T)
+\sum_{\alpha} S_{\mathrm{v}}^{\alpha}(t,T) w_{\alpha}(t,T).
\end{eqnarray}
Here $S_{\mathrm{v}}^{\alpha}(t,T)$ is the vibrational entropy at the basin
$\alpha$ such that $S_{\mathrm{v}}^{\alpha}(t,T)\equiv -k_{\mathrm{B}}
\int_{\Gamma_{\alpha}} w_{\mathrm{v}}^{\alpha}(\Gamma, t,T) \log
w_{\mathrm{v}}^{\alpha}(\Gamma, t,T)$.  
We add that
so-called configuration entropy 
is not necessarily equal to $S_{\mathrm{IS}}$ even in canonical equilibrium, 
$S_{\mathrm{IS,can}}(T) \neq k_{\mathrm{B}}\log \Omega_{\mathrm{IS}}{\small(U_{\mathrm{IS,can}}(T)\small)}$,
because
the value of $e_{\alpha}$ can largely deviate from
the mean value $U_{\mathrm{IS,can}}(T)$ in small systems such as
proteins. 
Here $\Omega_{\mathrm{IS}}(e_{\alpha})$ is the density of states 
for the inherent structures,
which can be determined 
from a numerical simulation in equilibrium situation
as shown in \cite{Nakagawa_Peyrard}.
In the present model $\Omega_{\mathrm{IS}}$ is almost 
an exponential function of $e_{\alpha}$,
similarly to simple models of glasses \cite{Bray,Doye,Fielding}.

\begin{figure}[t]
\includegraphics[width=9.5cm]{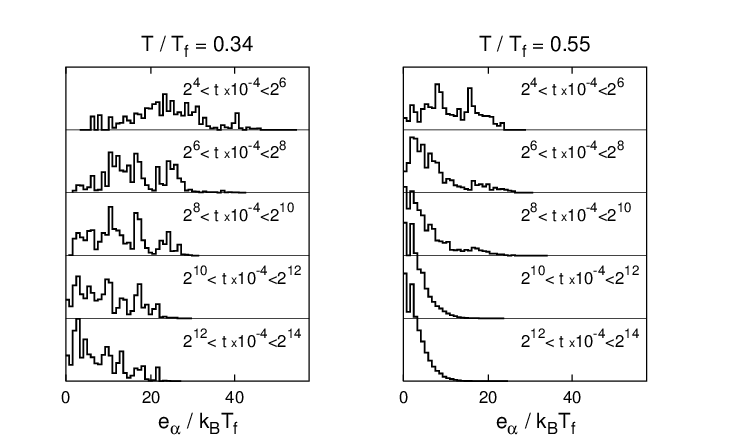}
\caption{
Time evolution of the probability distribution $P(e_{\alpha},t,T)$ 
for the inherent structure. 
The horizontal axis is inherent structure energy $e_{\alpha}$ scaled by
$k_{\mathrm{B}} T_f$.
$T=0.34 T_f$ and $0.55 T_f$ correspond to $T<T_d$ and $T>T_d$, respectively.
The right bottom figure shows the equilibrium distribution 
at $T=0.55T_f$. At $t=0$ the protein has been cooled to $T$.
}
\label{fig:P_IS}
\end{figure}

In analogy to the equilibrium case, we define a time-dependent 
conformational temperature for inherent structures as
\begin{eqnarray}
\frac{1}{T_{\mathrm{cnf}}(t,T)}\equiv
\frac{\partial S_{\mathrm{IS}}(t,T)/\partial T}{\partial U_{\mathrm{IS}}(t,T)/\partial T},
\label{eqn:effectiveT}
\end{eqnarray}
where we expect 
$T_{\mathrm{cnf}}(t, T) {\rightarrow} T$ as the relaxation proceeds, and $T_{\mathrm{cnf}}=T$ in thermal equilibrium.
Since $S_{\mathrm{IS}}$ and $U_{\mathrm{IS}}$ are a function of $t$ and $T$,
one may consider that 
$\frac{\partial S_{\mathrm{IS}}/\partial t}{\partial U_{\mathrm{IS}}/\partial t}$ 
can also give a temperature. 
In fact, the latter corresponds to the internal temperature defined in \cite{Sciortino}.
We have not adopted this definition because in the present protein system
$\frac{\partial S_{\mathrm{IS}}/\partial t}{\partial U_{\mathrm{IS}}/\partial t}$ 
does not depend on time
comparable to the change of the distribution seen in Fig. \ref{fig:P_IS}.
It is observed to remain in an approximately constant value 
corresponding to the exponent of $\Omega_{\mathrm{IS}}(e_{\alpha})$.

\paragraph*{Slow relaxation in Inherent structure landscape: }

\begin{figure}[t]
\includegraphics[width=7.5cm]{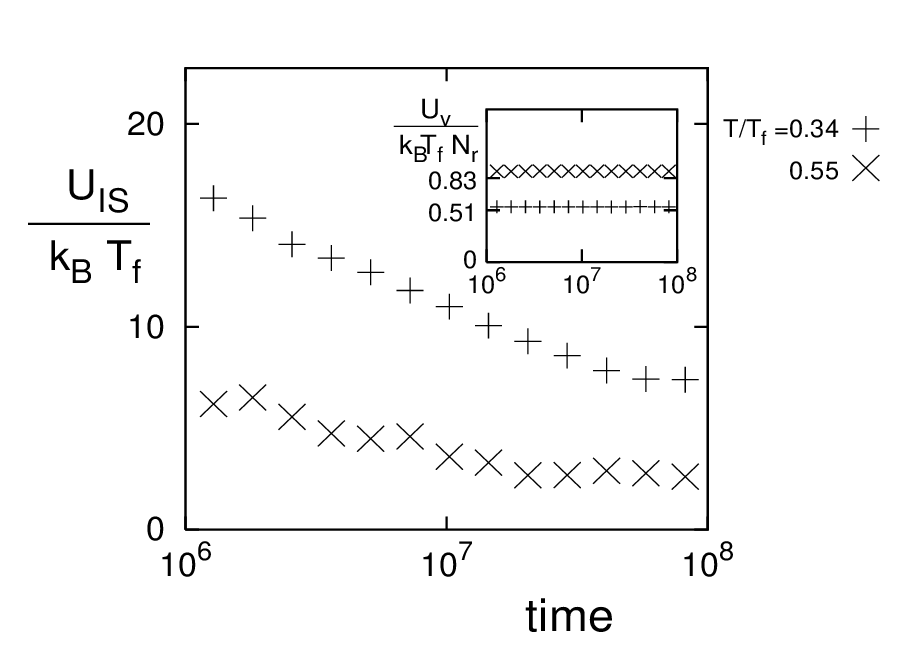}
\caption{
Time evolution of mean inherent structure energy $U_{\mathrm{IS}}$ and 
(inset) mean vibrational energy $U_{\mathrm{v}}$.  
In the case $0.55T_f> T_d$, the folding process reaches thermal equilibrium 
around $t=2\times 10^7$, 
whereas in $0.34T_f<T_d$ it does not within the computation time.
$U_{\mathrm{v}}$ is in the equilibrium value which is slightly larger 
than the harmonic approximation $3/2 k_{\mathrm{B}} N_r T$ (indicated
as $0.51$ and $0.83$ for $T/T_f=0.34$ and $0.55$).
$N_r$ is the number of residues.}
\label{fig:U_IS}
\end{figure}

The energetic aspects of the folding process are demonstrated in
Fig.\ref{fig:U_IS}, in which the mean vibrational energy over the phase
space is defined as $U_{\mathrm{v}}(t,T)\equiv
\sum_{\alpha}U_{\mathrm{v}}^{\alpha}(t,T)w_{\alpha}(t,T)$.  
As are shown in Fig.\ref{fig:U_IS}, the mean
inherent structure energy $U_{\mathrm{IS}}(t,T)$ relaxes logarithmically in
time. 
Such logarithmic
relaxations are typically seen in glassy systems, which suggests
intrinsic glassy properties for proteins even above the dynamical
  transition temperature $T_d$.

On the other hand,
$U_{\mathrm{v}}(t,T)$ is almost constant, corresponding to its equilibrium value
at temperature $T$,  
although
the probability weight $w_{\alpha}(t,T)$ is expected to significantly
depend on time as the folding proceeds. This is possible if the
vibrational energy $U_{\mathrm{v}}^{\alpha}(t,T)$ is approximately the same for
all basins $\alpha$ and  
the vibrational motions reach equilibrium much faster than the time
scale of the folding. 
The
value of $S_{\mathrm{v}}^{\alpha}(t,T)$ is also expected to be equal to its equilibrium value,
which could be approximately independent of $\alpha$, similarly to $U_{\mathrm{v}}^{\alpha}(t,T)$.

$S_{\mathrm{IS}}(t,T)$ cannot be accessible to numerical simulation
because the number of basins in the whole phase space is too large
to determine $w_{\alpha}(t,T)$ for each $\alpha$ and each $t$.
Instead of determining $w_{\alpha}(t,T)$ for all basins, 
let us introduce a probability $P(e_{\alpha},t,T)$ to observe 
the inherent structure energy $e_{\alpha}$, 
\begin{equation}
P(e_{\alpha},t,T)=\int d\Gamma w(\Gamma,t,T) 
\delta(e_{\mathrm{min}}(\Gamma)-e_{\alpha}).
\end{equation}
Here $e_{\mathrm{min}}(\Gamma)$ is the quenched energy from the point $\Gamma$.
If $\Gamma$ belongs to the basin $\alpha$, $e_{\mathrm{min}}(\Gamma)=e_{\alpha}$.

Since the initial conditions in numerical simulation are randomly selected 
from the random coil state,
let us assume that
the weight is approximately the same for basins with the same minimum energy 
$e_{\alpha}$,
\begin{equation}
\label{eq:approxw}
w_{\alpha}(t,T)\simeq P(e_{\alpha},t,T)/
\Omega_{\mathrm{IS}}(e_{\alpha}). 
\end{equation}
Then,
\begin{eqnarray}
S_{\mathrm{IS}}(t,T)\!\! &\simeq\! 
-k_{\mathrm{B}} \int\!  de_{\alpha} P(e_{\alpha},t,T) 
\log \frac{P(e_{\alpha}, t,T)}{\Omega_{\mathrm{IS}}(e_{\alpha})},
\label{eqn:Sis}
\end{eqnarray}
which is accessible to numerical simulations.

Examples of the time evolution of 
$P(e_{\alpha},t,T)$ in Fig.\ref{fig:P_IS} show that
the distribution is strongly dependent on time after the sudden cooling.
The energy range that the distribution spans
becomes smaller as time proceeds, and the width of the
distribution decreases significantly.
As is expected $P(e_{\alpha},t,T)$ is considerably different from 
the equilibrium distribution (bottom figure for $T/T_f=0.55$).

\begin{figure}[t]
\includegraphics[width=6.cm]{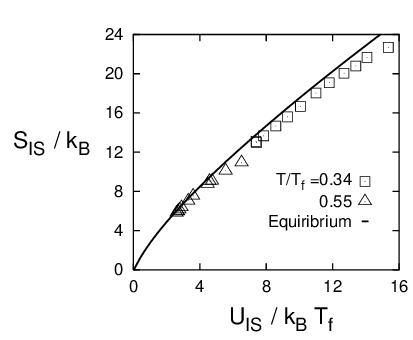}
\caption{
Time evolution in the plot on $S_{\mathrm{IS}}$ vs $U_{\mathrm{IS}}$
for $t>10^6$.
$T/T_f=0.34$ and $0.55$.
The line is the relation between $S_{\mathrm{IS}}$ and $U_{\mathrm{IS}}$ in thermal equilibrium.
Higher temperatures corresponds to the right-upper part of the line. 
}
\label{fig:Sis-Uis}
\end{figure}

\begin{figure}[t]
\includegraphics[width=9cm]{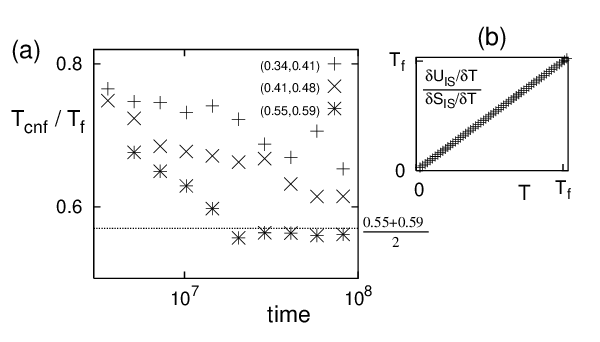}
\caption{
(a) Time evolution of $T_{\mathrm{cnf}}(t, (T_1+T_2)/2)$ in eq.(\ref{eqn:effectiveT}) 
estimated by
$\{S_{\mathrm{IS}}(t,T_2)-S_{\mathrm{IS}}(t,T_1)\}/(T_2-T_1)$ and
$\{U_{\mathrm{IS}}(t,T_2)-U_{\mathrm{IS}}(t,T_1)\}/(T_2-T_1)$.
Here $(T_1, T_2)=(0.34T_f, 0.41T_f), 
(0.41T_f, 0.48T_f)$ and $(0.55T_f, 0.59T_f)$.
The dotted horizontal line is $(T_1+T_2)/2$ for $(0.55T_f, 0.59T_f)$,
which is expected to be a convergence line of $T_{\mathrm{cnf}}(t,0.57T_f)$.
(b) Confirmation of thermodynamic 
relation, $T_{\mathrm{cnf}}=T$ in thermal equilibrium condition.
}
\label{fig:effectiveT}
\end{figure}

\begin{figure}[t]
\includegraphics[width=6cm]{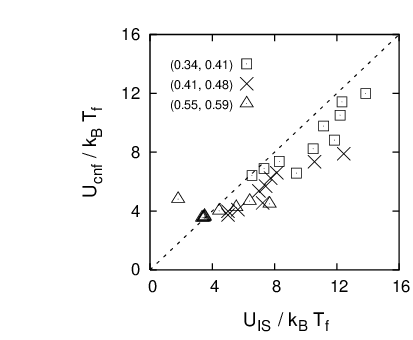}
\caption{
(a) $U_{\mathrm{cnf}}(t, T)$ vs $U_{\mathrm{IS}}(t, T)$.
Here $T=(T_1+T_2)/2$ with $(T_1, T_2)=(0.34T_f, 0.41T_f), 
(0.41T_f, 0.48T_f)$ and $(0.55T_f, 0.59T_f)$.
}
\label{fig:Ue-Uis}
\end{figure}

For each time $t$, we can obtain numerically $U_{\mathrm{IS}}(t,T)$ and 
$S_{\mathrm{IS}}(t,T)$ following the definition (\ref{eqn:Sis}).
If we plot $S_{\mathrm{IS}}(t,T)$ as a function of $U_{\mathrm{IS}}(t,T)$ for various $t$,
we get the time evolution for the relaxation process 
shown in Fig.\ref{fig:Sis-Uis}.
For comparison we plot the line corresponding to 
thermal equilibrium cases over a wide range of $T$.
This comparison makes it clear that the time evolution 
in a fixed cooling temperature $T$
proceeds in parallel to the line for 
thermal equilibrium from high to low temperatures.
Let us study this evolution by
the conformational temperature $T_{\mathrm{cnf}}(t,T)$ defined in eq.~(\ref{eqn:effectiveT}).
In equilibrium, $S_{\mathrm{IS}}$ and $U_{\mathrm{IS}}$ 
work well to determine the temperature of the system
as is demonstrated in Fig.~\ref{fig:effectiveT}(b),
where $T_{\mathrm{cnf}}$ for the canonical ensemble with $\Omega_{\mathrm{IS}}(e_{\alpha})$
has the expected property, $T_{\mathrm{cnf}}=T$.
In the relaxation process, 
$T_{\mathrm{cnf}}$ 
decreases with time from a higher temperature.
When $T$ is sufficiently high, we can observe that $T_{\mathrm{cnf}}$ converges
to a value corresponding to $T$ 
after the relaxation to equilibrium
($T\simeq 0.55T_f$ in Fig.~\ref{fig:effectiveT}(a)).
Thus 
the features observed in Fig.~\ref{fig:P_IS} or Fig.~\ref{fig:Sis-Uis}
in the relaxation process are reflected in $T_{\mathrm{cnf}}(t, T)$.

The value of $T_{\mathrm{cnf}}(t, T)$ is expected to give an approximation of $P(e_{\alpha}, t, T)$
to an equilibrium distribution at the temperature $T_{\mathrm{cnf}}$. 
To check it,
we compare the approximated mean IS-energy
$U_{\mathrm{cnf}}(t, T)$ with $U_{\mathrm{IS}}(t, T)$ in Fig.~\ref{fig:Ue-Uis}, where
\begin{equation}
U_{\mathrm{cnf}}(t, T) \equiv \int \! d{e_{\alpha}} 
\frac{e_{\alpha} ~\Omega_{\mathrm{IS}}(e_{\alpha})}{Z_{\mathrm{IS}}(T_{\mathrm{cnf}}(t, T))} 
\exp\left(\frac{-e_{\alpha}}{k_B T_{\mathrm{cnf}}(t, T)}\right).
\label{eqn:U_e}
\end{equation}
The figure displays the tendency $U_{\mathrm{cnf}}(t, T)\simeq U_{\mathrm{IS}}(t, T)$, which suggests that 
$T_{\mathrm{cnf}}$ can give a first approximation for the distribution $P(e_{\alpha}, t, T)$.
The deviation of the points from the line 
in Fig.~\ref{fig:Ue-Uis} indicates that
the states on the way of folding
do not coincide strictly with any equilibrium state.
Correspondingly, the points in Fig.~\ref{fig:Sis-Uis} deviates slightly
from the equilibrium line.

\paragraph*{Discussion:}

We have introduced time dependent energies and entropies for inherent
structures 
to analyze the slow folding process of a model protein.
After a sudden cooling, we found that the probability distribution of the 
inherent structures 
is widely spread in the early stage and becomes narrower as time proceeds.
This evolution is successfully characterized by the time evolution of
the conformational temperature Eq.(\ref{eqn:effectiveT}), 
as shown in Fig.\ref{fig:effectiveT}(a).

Although the approximation of the distribution by $T_{\mathrm{cnf}}$ is not complete, 
we note that 
even this first approximation is not obtained if we treat the
total probability distribution without separating the inherent structure modes 
and the vibrational modes.
As argued in Fig.\ref{fig:U_IS}, the vibrational modes are reaching 
local equilibrium in the early stage.
Thus there is a temperature difference between the fast dynamics and
the slow dynamics. 
Thermodynamic frameworks including two temperatures 
could be proposed \cite{Nieuwenhuizen}.
It might be interesting to study various slow dynamics 
adopting such two temperatures.

Sciortino and Tartaglia \cite{Sciortino}
introduced an internal temperature for glasses 
as a differential of two temporal states characterized by
the inherent structure energy and configuration entropy.
The internal temperature appears to approximate temporal distribution 
$P(e_{\alpha}, t, T)$ in some glasses \cite{Crisanti-Ritort}
although we observe that {\it it does not in the protein model}.
From these studies, we expect that internal temperature  
coincides with confromational temperature in glassy systems 
in contrast with their disagreement in the protein model.
This difference might reprenst a defference in a basic property of the system, 
i.e., glasses are homogeneous and extensive but single proteins
 are heterogeneous and nonextensive.
The measurement of effective temperature in protein models,
which is defined by the violation of fluctuation-dissipation relation \cite{Cugliandolo1,Cugliandolo2}
and has known to coincide with the internal temperature in a Lenard-Jones or a spin glass \cite{Sciortino,Crisanti-Ritort},
 is a future work to be studied.
This is not a simple task because the effective temperature could be
observable-dependent \cite{Fielding} and the intensive variable
for the protein models is not obvious.

Finally we would like to point out that this study emphasizes the
interest of the Inherent Structure Landscape for protein. We showed
earlier \cite{Nakagawa_Peyrard} that, contrary to the free-energy
landscape which is only a useful concept, the density of states of the
ISL can be obtained in practice for a protein model, and that it
characterizes the equilibrium properties of the protein to a good
accuracy. Here we show that the ISL is not only useful in equilibrium,
but can be used to characterize the slow relaxation with a
time-dependent conformational temperature.
The assumption used in this letter, that
$U_{\mathrm{v}}^{\alpha}$ and $S_{\mathrm{v}}^{\alpha}$ does
not depend on $\alpha$, appears to be good for the protein case.
It can be relaxed for more complex systems \cite{Nakagawa_Peyrard-future}.

Note: Since the present model possesses at least two funnel which is well
separated from the others,  we choose the relaxation process to a funnel 
of which global minimum energy state 
corresponds to a misfolded state.  Its minimum
energy state resembles the native state but a
$\alpha$-helix is bent compared to the native state.
In this funnel, relaxation proceeds in a time scale accessible 
to numerical simulations if the temperature is not too low.

Acknowledgement: N.N. thanks M. Peyrard for useful discussion 
and critical reading of the manuscript. She also thanks MEXT KAKENHI (No. 16740217) for support.

\end{document}